\documentclass{article}
\usepackage{spconf,amsmath,graphicx}
\usepackage{comment}
\usepackage{multirow}
\usepackage{makecell}
\usepackage{caption}
\usepackage{url}
\usepackage{amsmath}
\def\*#1{\mathbf{#1}}
\usepackage[noadjust]{cite}
\usepackage{caption}
\usepackage{subcaption}
\usepackage[symbol]{footmisc}
\usepackage{cleveref}
\usepackage[flushleft]{threeparttable}
\usepackage{marvosym}

\title{Robust speaker recognition using unsupervised adversarial invariance}

\name{Raghuveer Peri, Monisankha Pal, Arindam Jati, Krishna Somandepalli, Shrikanth Narayanan}
\address{Signal Analysis and Interpretation Laboratory, University of Southern California, Los Angeles, CA, USA}
\begin{document}
\ninept
\maketitle
\begin{abstract}

In this paper, we address the problem of speaker recognition in challenging acoustic conditions using a novel method to extract robust speaker-discriminative speech representations. We adopt a recently proposed unsupervised adversarial invariance architecture to train a network that maps speaker embeddings extracted using a pre-trained model onto two lower dimensional embedding spaces. The embedding spaces are learnt to disentangle speaker-discriminative information from all other information present in the audio recordings, without supervision about the acoustic conditions. We analyze the robustness of the proposed embeddings to various sources of variability present in the signal for speaker verification and unsupervised clustering tasks on a large-scale speaker recognition corpus.
Our analyses show that the proposed system substantially outperforms the baseline in a variety of challenging acoustic scenarios. Furthermore, for the task of speaker diarization on a real-world meeting corpus, our system shows a relative improvement of 36\% in the diarization error rate compared to the state-of-the-art baseline.

\end{abstract}
\section{Introduction}
\label{sec:intro}

Obtaining robust \textit{speaker embeddings} i.e., low-dimensional representations from speech signals that capture speaker characteristics, is a particularly challenging problem given the diverse nature of possible variability in the signal.
The signal variability could arise from various nuisance factors such as background acoustic noise, room reverberation, microphone placement etc. The presence of such variability in the signal makes tasks which rely on speaker-discriminative features such as speaker verification and speaker diarization even more challenging \cite{sell2018diarization}. This serves as a motivation to extract speaker embeddings that are invariant to nuisance factors.

Until recently, much of the speaker verification research was based on generative modeling based embeddings such as i-vectors \cite{5545402}. 

Since i-vectors contain both speaker and channel information, they fail to provide speaker embeddings robust to the nuisance factors \cite{6288859}, requiring additional supervised compensation steps \cite{solomonoff2007nuisance,Kanagasundaram2011ivectorBS}.
With the advances in deep learning technologies, robust speaker modeling approaches based on neural networks have been proposed
\cite{snyder2017deep,bhattacharya2017deep,snyder2018x}. These techniques can be broadly categorized into two classes: data augmentation and adversarial invariance.
In ~\cite{snyder2017deep}, a time-delay deep neural network~(TDNN) based model was proposed, that was trained on variable length utterances to generate fixed length speaker representations using a cross entropy loss.
In \cite{snyder2018x}, a large corpus of audio recordings from various sources was combined, which was further augmented by artificially adding background noise and music at varying signal-to-noise levels. In order to simulate the effect of reverberation, audio signal was convolved with various room impulse responses. Speaker embeddings, called \textit{x-vectors}, extracted using this technique have 
provided state-of-the-art performance in applications such as speaker verification \cite{snyder2018x} and diarization \cite{7953094}.
One inherent drawback with such data augmentation approaches is that they learn specific variations of the acoustic signal and tend to degrade in performance when tested on unseen acoustic variations, as shown in \cite{Jati_IS}. Further, data augmentation techniques do not explicitly ensure that irrelevant information is removed from the speaker representations, as shown through various probing tasks in \cite{raj2019probing}.

A promising research direction in this context is domain adversarial training to make speaker representations robust to recording conditions \cite{8682064,zhou2019training,meng2019adversarial,tu2019variational}. However, a majority of these techniques are supervised, i.e., they require labelled nuisance factors, which might not be readily available in many real-world scenarios.
This necessitates unsupervised adversarial training, that can learn speaker representations robust to channel and other acoustic variability without knowledge of any particular nuisance factor. Such work in the
speech domain has been largely unexplored. Recently, an unsupervised approach to induce invariance for automatic speech recognition was introduced in \cite{Hsu2019NIESRNI}. However, the goal of that work was to remove speaker-specific information from the speech representations.

In this work, we explore a method of inducing robustness in speaker embeddings to cope with challenging acoustic environments, where no prior information about the recording conditions is readily available.
We adopt a recently proposed unsupervised adversarial invariance ~(UAI) architecture \cite{jaiswal2018unsupervised} to extract two disentangled representations from x-vectors. One representation is trained to contain only speaker-discriminative information, while all other information is captured in the second embedding.
We empirically show that embedding learnt using our method is able to capture speaker-related information better than the decoderinput x-vectors, while forcing other information pertaining to the nuisance factors to be captured in a second embedding.
Our proposed method is different from previously proposed adversarial invariance techniques for speaker embeddings in that, our model does not rely on any supervised information about the nuisance factors.

\vspace{-4pt}
\section{METHODS}
\label{sec:methods}

\subsection{Feature extraction}
\label{ssec:x-vector}
As described before, x-vectors have shown to be robust for speaker recognition tasks and achieve state-of-the-art performance. Therefore, we use these as input features for our model.
The input features were extracted using a pre-trained, publicly available TDNN based embedding system\footnote{\url{https://kaldi-asr.org/models/m7}}. It takes frame-level MFCC features as input and produces segment-level x-vectors. 

\subsection{Adversarial nuisance invariance}
\label{ssec:UAI}
Although x-vectors have produced state-of-the-art performance for speaker recognition tasks, they have also been shown to capture information related to nuisance factors \cite{raj2019probing}. Our objective, in using unsupevised invariance technique, is to further remove the effects of the nuisance signals from the x-vectors.

\begin{figure}[!t]
  \centering
  \centerline{\includegraphics[width=9cm]{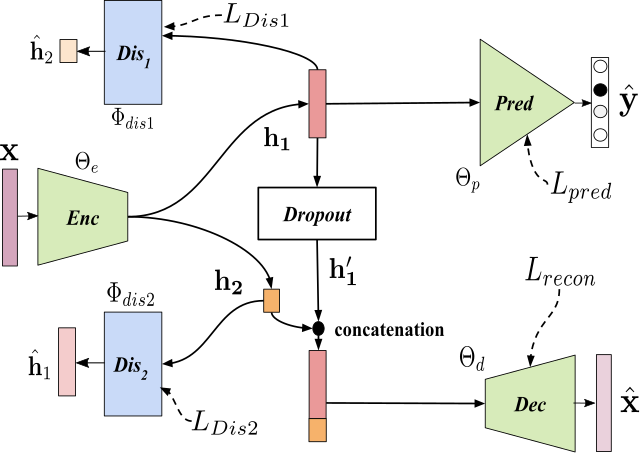}}
\caption{Unsupervised adversarial invariance applied for speaker recognition}
\label{fig:uai}
\end{figure}

Fig.~\ref{fig:uai} shows the full UAI architecture, which consists of an encoder (\textit{Enc}), decoder (\textit{Dec}), predictor (\textit{Pred}) and two disentanglers ($\textit{Dis}_{1}$) and ($\textit{Dis}_{2}$). \textit{Enc} maps the input utterance-level x-vector $\mathbf{x}$ into two latent representations $\mathbf{h}_{1}$ and $\mathbf{h}_{2}$, each used for different downstream tasks. \textit{Pred} classifies $\mathbf{h}_{1}$ as belonging to one of the known speakers producing a one-hot encoded representation at its output, $\hat{\mathbf{y}}$. Meanwhile, a dropout module, \textit{Dropout}, randomly removes some dimensions from $\mathbf{h}_{1}$ to create a noisy version denoted by $\mathbf{h}_{1}^\prime$. Then the decoder \textit{Dec} takes a concatenation of $\mathbf{h}_{1}^\prime$ and $\mathbf{h}_{2}$ and reconstructs the input x-vector, denoted by $\hat{\mathbf{x}}$. In addition, the latent embeddings $\mathbf{h}_{1}$ and $\mathbf{h}_{2}$ are passed through two different modules $\textit{Dis}_{1}$ and $\textit{Dis}_{2}$. The goal of these modules is to predict $\mathbf{h}_{2}$ from $\mathbf{h}_{1}$ and vice-versa. The parameters of encoder, decoder and predictor are denoted by $\Theta_{e}$, $\Theta_{d}$ and $\Theta_{p}$ respectively, while those of the disentanglers are denoted by $\Phi_{dis1}$, $\Phi_{dis2}$. We use categorical cross entropy loss for the predictor ($L_{pred}$) and mean square error loss for both the disentanglers as well as for the decoder ($L_{recon}$, $L_{Dis1}$, $L_{Dis2}$)

The central idea in the UAI method is to setup a minimax game between the main model comprising the modules \textit{Enc}, \textit{Dec} and \textit{Pred} and the adversarial model comprising $\textit{Dis}_{1}$ and $\textit{Dis}_{2}$. The goal here is to maximize the predictive power of $\mathbf{h}_{1}$ for speaker classification and reconstruct the input features with $\mathbf{h}_{2}$, simultaneously minimizing the predictive power between $\mathbf{h}_{1}$ and $\mathbf{h}_{2}$, thereby disentangling the two representations. Equations \eqref{eq1} and \eqref{eq2} describe the loss functions of the main and adversarial models, respectively. 

\begin{equation}
L_{main} = \alpha L_{pred}\left(\mathbf{y},\hat{\mathbf{y}}\right) + \beta L_{recon}\left(\mathbf{x},\hat{\mathbf{x}}\right) 
\label{eq1}
\end{equation}

\begin{equation}
L_{adv} = L_{Dis1}(\mathbf{h}_{2},\hat{\mathbf{h}}_{2}) + L_{Dis2}(\mathbf{h}_{1},\hat{\mathbf{h}}_{1}) 
\label{eq2}
\end{equation}

The task of the decoder is to reconstruct the input x-vectors by minimizing $L_{recon}$. Since the decoder receives a noisy version of $\mathbf{h}_{1}$, as the training progresses, \textit{Dec} learns to treat $\mathbf{h}_{1}$ as an unreliable source of information for the reconstruction task and thus is forced to squeeze all information into $\mathbf{h}_{2}$. However, this is not sufficient to ensure that $\mathbf{h}_{1}$ and $\mathbf{h}_{2}$ contain complementary information. Hence, explicit "disentanglement" between the two latent representations is encouraged by training the two tasks involving $L_{main}$ and $L_{adv}$ in an adversarial fashion, consistent with related previous work \cite{jaiswal2018unsupervised, xie2017controllable}.
The adversarial training can be setup as shown in Equation \eqref{eq:L_all}, where $\alpha$, $\beta$ and $\gamma$ are tunable parameters.

\begin{equation}
\label{eq:L_all}
    \underset{\Theta_{e},\Theta_{d},\Theta_{p}}{\textnormal{min}} \hspace{1pt} \underset{\Phi_{dis1}, \Phi_{dis2}}{\textnormal{max}} L_{main} +\\
    \gamma L_{adv}
\end{equation}

In our experiments, the modules, \textit{Enc}, \textit{Dec}, \textit{Pred}, $\textit{Dis}_{1}$ and $\textit{Dis}_{2}$ comprised of $2$ hidden layers each. \textit{Enc} and \textit{Dec} had $512$ units in each layer, while the disentanger modules had $128$ units in each layer. For \textit{Pred}, $256$ and $512$ were used as number of hidden units. The dropout probability of the \textit{Dropout} module was set to $0.75$. We set the weights for the losses as $\alpha = 100$, $\beta = 5$ and $\gamma = 50$. Parameters were tuned by observing the convergence of the losses on a pre-determined subset of the training data. The model was trained for $350$ epochs with a batch size of $128$. In each epoch, the model was trained with $10$ batches of adversarial model update for every $1$ batch of main model update. Both objectives were optimized using the Adam optimizer with $1e-3$ and $1e-4$ learning rates respectively and a weight decay factor of $1e-4$ for both. The dimensions for the embeddings $\mathbf{h}_{1}$ and $\mathbf{h}_{2}$ were chosen as $128$ and $32$ respectively.

\section{DATASETS}
\label{sec:data}
\begin{figure}[t]
  \centering
  \centerline{\includegraphics[width=6cm,height=3.5cm]{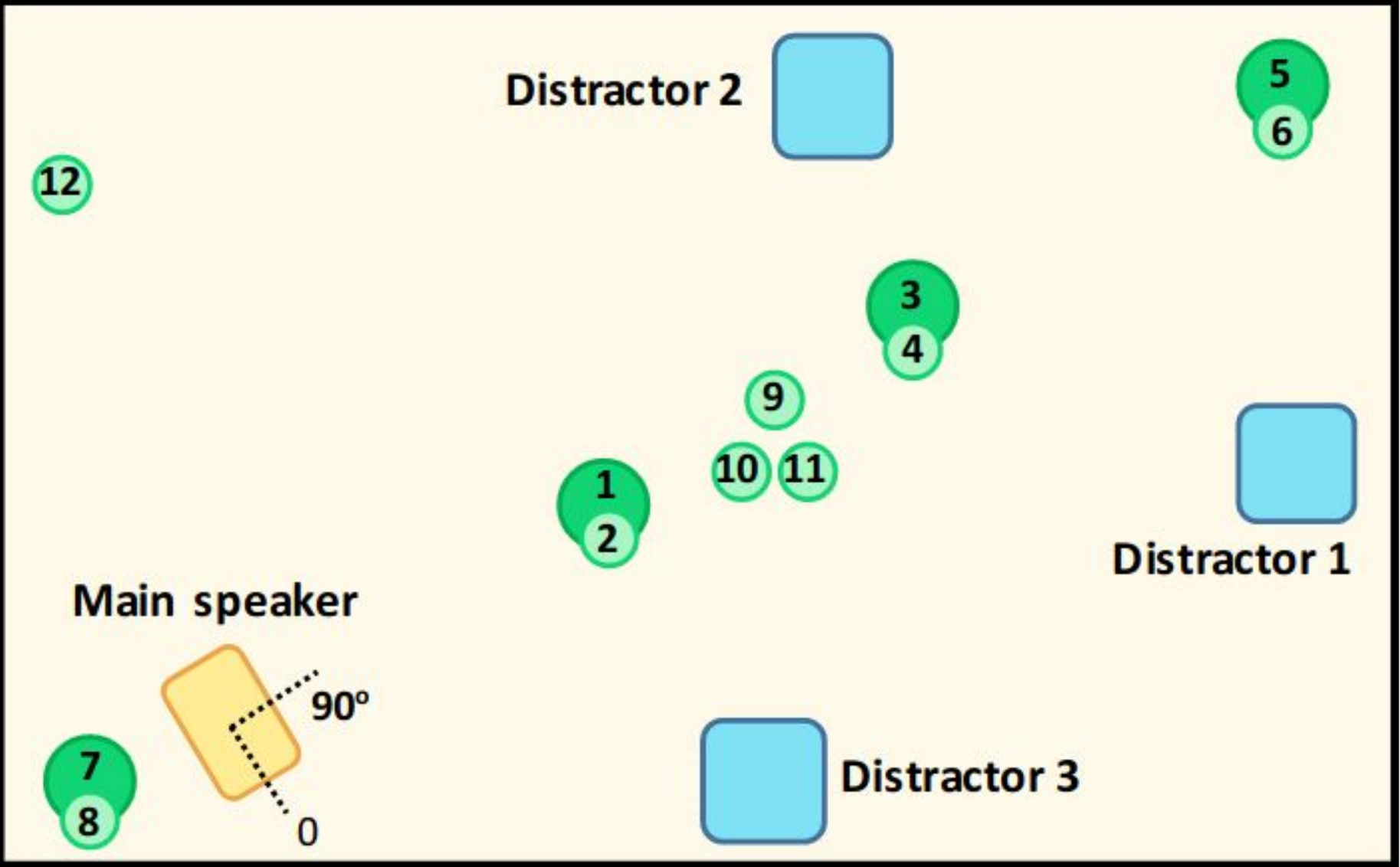}}

\caption{Example room configuration in VOiCES dataset \cite{richey2018voices}. Distractor represents noise source and green circles represents microphones}
\label{fig:voices}
\end{figure}

In this work, we designed experiments to analyze general speaker verification performance, while also performing controlled experiments for two main sources of variability that can occur in real-world audio recordings. We also perform probing tasks to understand better the information contained in the speaker embeddings that we extract. In this section, we provide details of the publicly available datasets that we use for the experiments.
\vspace{5pt} \\
\textbf{{AMI}}:
To evaluate the performance of the proposed embeddings on the speaker diarization task, we use a subset of the AMI meeting corpus \cite{mccowan2005ami} that is frequently used for evaluating diarization performance \cite{sun2019speaker, cyrta2017speaker}. It consists of audio recordings from $26$ meetings.
\vspace{5pt} \\
\textbf{{V19-eval}} and \textbf{{V19-dev}}: We use the VOiCES data corpus \cite{richey2018voices} to evaluate the performance of our system with respect to the baselines on a speaker verification task and perform probing tasks to examine the systems. It consists of recordings collected from $4$ different rooms with microphones placed at various fixed locations, while a loudspeaker played clean speech samples from the Librispeech \cite{panayotov2015librispeech} dataset. Along with speech, noise was played back from loudspeakers present in the room, to simulate real-life recording conditions. Fig~\ref{fig:voices} shows one such room configuration and data collection setup where "Distractor" represents noise source and the green circles represent the $12$ available microphones.

We use two subsets of this data corpus, the development portion of the VOiCES challenge data \cite{nandwana2019voices} referred to as V19-dev and the evaluation portion referred to as V19-eval. V19-dev is used for probing experiments as discussed in Section \ref{ssec:clustering}, as it contains annotations for $200$ speaker labels, $12$ microphone locations and $4$ noise types ~(none, babble, television, music). V19-eval is used for experiments to study the robustness of the systems to various nuisance factors. We use the evaluation portion of the dataset for robustness analysis as it contained more challenging recording conditions than the development portion.

\vspace{5pt}
\textbf{{Vox}}:
Our training data consists of a combination of the development and test splits of VoxCeleb2 \cite{chung2018voxceleb2} and the development split of VoxCeleb1 \cite{nagrani2017voxceleb} datasets. This is consistent with the split that was used to train the pre-trained x-vector model (mentioned in Section \ref{ssec:x-vector}), but with no data augmentation. It consists of speaker annotated \textit{in-the-wild} recordings from celebrity speakers. As such the dataset is sourced from unconstrained recording conditions. For brevity, henceforth, we refer to this subset of the VoxCeleb dataset as Vox.
\vspace{5pt} \\
Table \ref{table:data} shows the statistics for the different datasets used in our work. We ensured that the speakers contained in one dataset had no overlap with the speakers from any other dataset.

\begin{table}[!t]
\caption{Statistics of datasets (utt refers to utterances, spk refers to speakers)}
\label{table:data}
\begin{threeparttable}
\begin{tabular}{ccccc}

\Xhline{2.5\arrayrulewidth}
Name              & Purpose     & No.utt  & No.spk & \begin{tabular}[c]{@{}c@{}}Nuisance \\ annotations\\ available\end{tabular} \\ \Xhline{2.5\arrayrulewidth}
AMI      & diarization &          26\tnote{\Cross}  &  29  & no                                                                          \\ 
V19-eval & verification     &          11,392 &  47    & yes                                                                         \\ 
V19-dev  & clustering       & 15,904 &  200  & yes                                                                         \\ 

Vox      & train       & 1.2M    & 7323  & no                                                                          \\ \Xhline{2.5\arrayrulewidth}

\end{tabular}
\begin{tablenotes}
        \item[\Cross] \textit{Refers to number of sessions}
\end{tablenotes}
\end{threeparttable}
\end{table}

\section{EXPERIMENTS}
\label{sec:experiments}

We setup the following experiments to study the different aspects of our system:
\begin{enumerate}
    \item Robustness analysis of speaker verification (V19-eval dataset)
    \item Unsupervised clustering (V19-dev dataset)
    \item Speaker diarization with oracle speech segment boundaries (AMI dataset)
\end{enumerate}
\textbf{Baseline}: We used x-vectors extracted from the pre-trained model\footnote{\url{https://kaldi-asr.org/models/m7}} as baseline, to test if the proposed model is able to improve robustness of speaker embeddings by removing the nuisance factors from x-vectors. In the results in Sections \ref{ssec:SV} and \ref{ssec:clustering}, we denote the baseline method by x-vector and the method using the proposed embeddings by $\mathbf{h_{1}}$. In Section \ref{ssec:diar}, the baselines are denoted by Baseline 1 and Baseline 2, which are defined in the section.

\begin{table}[!t]
\caption{Speaker verification (\% EER) vs. nuisance factors (V19-eval)}
\label{table:voices_analysis}
\setlength{\tabcolsep}{11pt}
\centering
\begin{tabular}{cc|cc}
\Xhline{3.5\arrayrulewidth}
\multicolumn{2}{c|}{}                                                            & x-vector      & $\mathbf{h_{1}}$        \\ \Xhline{2.5\arrayrulewidth}

\multirow{3}{*}{\begin{tabular}[c]{@{}c@{}}noise\\ (near-mic)\end{tabular}} & none & \textbf{3.34} & 3.99           \\ 
                                                                            & babble  & 5.41          & \textbf{4.86}  \\ 
                                                                            & television  & \textbf{3.28} & 4.15           \\ \hline
                                                                            
\multirow{3}{*}{\begin{tabular}[c]{@{}c@{}}noise \\ (far-mic)\end{tabular}} & none & 7.43          & \textbf{6.26}  \\ 
                                                                            & babble  & 21.93         & \textbf{19.79} \\ 
                                                                            & television  & 10.80         & \textbf{9.05}  \\ \Xhline{2.5\arrayrulewidth}

\multirow{3}{*}{\begin{tabular}[c]{@{}c@{}}mic \\ placement\end{tabular}}   & near-mic  & \textbf{4.17} & 4.41           \\

& far-mic   & 14.97         & \textbf{12.79} \\ 
                                                                            & obstructed-mic  & 6.34          & \textbf{5.67}  \\ 
                                                                             \Xhline{2.5\arrayrulewidth} 
\multicolumn{2}{c|}{overall}                                                        & 10.30         & \textbf{9.07}  \\ \Xhline{3.5\arrayrulewidth}
\end{tabular}
\end{table}

\subsection{Speaker Verification}
\label{ssec:SV}

\subsubsection{Setup}
We evaluate the baseline and the proposed methods for verification task on the V19-eval dataset. Following standard practice \cite{snyder2018x}, we perform dimensionality reduction using linear discriminant analysis~(LDA) and score the verification trials using a probabilistic linear discriminant analysis~(PLDA) backend for both the proposed embeddings and the baseline. The LDA and PLDA models were learnt on the training data for our proposed system, while for the baseline system we used the pre-trained models. For the embeddings extracted using our method, we use a dimension of $96$ after LDA, while for x-vectors we use $150$ as the reduced dimension. Consistent with general practice \cite{snyder2017deep}, equal error rate~(EER) was used as the metric for evaluation.

Following \cite{richey2018voices} and \cite{nandwana2018robust}, we used knowledge of the nuisance factors annotations available in V19-eval dataset to study the various factors affecting the performance of speaker verification. For these experiments we consider two distinct nuisance factors, noise conditions: none, babble and television and microphone location: far-mic, near-mic and obstructed-mic (microphone hidden in the ceiling).

We further distinguish between the recordings collected at $2$ different microphone locations (far-mic vs. near-mic) while examining the performance in noisy conditions.

The experimental setup for the controlled conditions is shown in Fig~\ref{fig:voices}. As mentioned in Section \ref{sec:data}, the green circles, numbered 1-12, represent microphones located at various distances from the main loudspeaker. In all experiments, the enrolment utterances were collected from source data used to playback from the loudspeaker, consistent with \cite{nandwana2019voices}. We choose a different set of test utterances depending on the experiment being performed. For example, to evaluate performance in the noisy (near-mic) scenario, we use the utterances that were recorded from mics 1 and 2 as shown in Fig.~\ref{fig:voices} as the test utterances. Similarly for the noisy (far-mic) scenario, test utterances are pooled from mics 5 and 6.

\subsubsection{Results}
As shown in Table \ref{table:voices_analysis}, from the analysis on the effect of noise, although the baseline provides better performance in near-mic scenario for no noise and television noise conditions, the verification performance using the proposed embedding (denoted by $\mathbf{h_{1}}$ in Table \ref{table:voices_analysis}) provides improvement over the baseline when the test utterances were recorded at distant microphones, for all the noise types. As previously observed in \cite{nandwana2018robust}, babble noise seems to be the most challenging of all the noise types in terms of verification performance due to its speech-like characteristics. In this particularly harsh condition, the proposed embedding outperforms the baseline in both the near-mic and far-mic scenarios. Interestingly, our method shows the highest absolute improvement ($\sim{2.2}\%$ in EER) in the most challenging condition, i.e., far-mic recording in the presence of babble noise.

In experiments on the effect of microphone placement, the results show that our method performs comparable to the baseline in near-mic scenario and outperforms the baseline in the more challenging far-mic and obstructed-mic scenarios.

The last row in Table \ref{table:voices_analysis} shows the overall speaker verification performance using test utterances from all the microphones under all noise conditions. In this experiment we see a relative 10\% EER improvement by our system over the baseline.

\subsection{Clustering analysis of embeddings}
\label{ssec:clustering}
\subsubsection{Setup}
In order to further probe the information contained in the latent representations, we analyze clustering performance of the embeddings. 
We expect $\mathbf{h_{1}}$ to perform best when clustering speakers and $\mathbf{h_{2}}$ to cluster well with respect to the nuisance factors.
We use normalized mutual information~(NMI) between the embeddings and the ground truth clusters as a proxy for the speaker/nuisance related information contained in each of the embeddings. The ground truth clusters here are obtained from the annotations available in the V19-dev dataset. Clustering is performed using k-means (mini batch k-means implementation in \cite{scikit-learn}) with the known number of clusters.

\subsubsection{Results}
\renewcommand{\thefootnote}{\fnsymbol{footnote}}
Table \ref{table:cluster_analysis} reports results comparing the performance of both our embeddings and the baseline in clustering speakers and nuisance factors (noise type and microphone location).
We conducted permutation tests \cite{raschkas_2018_mlxtend} between the clustering results of the different experiments\footnote{Reject null hypothesis that the results come from same ditribution if \textit{p}-value $<\alpha$ where $\alpha=0.025$ to account for multiple comparison testing} to test for statistical significance.

Clustering by speaker, we see that $\mathbf{h_{1}}$ performs significantly better than x-vectors (absolute 4.3\% as shown in Table \ref{table:cluster_analysis}). This suggests that our method is able to extract more speaker-discriminative information from x-vectors. Furthermore, as expected, $\mathbf{h_{2}}$ showed relatively poor performance in clustering speakers.

Clustering by nuisance factors, as expected, $\mathbf{h_{2}}$ is the most predictive. Also, $\mathbf{h_{1}}$ doesn't cluster well according to the nuisance factors. Consistent with our findings in Section \ref{ssec:SV}, x-vectors have a significantly higher NMI scores than $\mathbf{h_{1}}$ (row 3, Table \ref{table:cluster_analysis})). This suggests that the proposed embedding is able to capture speaker information, invariant to microphone placement better than x-vectors. We found significant differences in the reported NMI scores, suggesting that our method is able to disentangle the $2$ different streams of information, speaker-related and nuisance-related.

\subsection{Speaker diarization using oracle speech segment boundaries}
\label{ssec:diar}
We further extend the analysis by examining the effectiveness of our proposed speaker embedding in speaker diarization task \cite{garcia2017speaker, sell2018diarization}. Since the goal of this work is to investigate the speaker-discriminative nature of embeddings, we consider only speaker clustering in the diarization task and assume prior knowledge of speaker homogeneous segments and the number of speakers, as was done in past studies \cite{garcia2017speaker, sun2019speaker}. The proposed speaker diarization system (denoted by $\mathbf{h_{1}}$) is based on embeddings extracted from speaker-homogeneous segments followed by k-means clustering as the backend. We compare our system with two competitive baselines that use x-vectors from pre-trained model as input features. One baseline (denoted by Baseline 1) uses k-means clustering on the extracted x-vectors. The other baseline is the state-of-the-art diarization system \cite{garcia2017speaker} (denoted by Baseline 2) which uses PLDA scoring and agglomerative hierarchical clustering~(AHC).

Diarization error rate (DER) \cite{fiscus2006rich} averaged across all sessions in the AMI dataset are shown in Table \ref{table:sd}. We see that our proposed system outperforms both Baseline 1 and Baseline 2 systems by a relative $38\%$ and $36\%$ in DER respectively. This suggests that the proposed speaker embeddings contain more speaker discriminative information than x-vector embeddings and hence are better suited for speaker clustering across datasets.

\begin{table}[!t]
\caption{Normalized mutual information (\%) between clusters of embeddings and true cluster labels. k represents no. clusters (V19-dev)}
\label{table:cluster_analysis}
\setlength{\tabcolsep}{8.5pt}
\centering
\begin{tabular}{c|ccc}
\Xhline{2.5\arrayrulewidth}
& $\mathbf{h_{1}}$        & $\mathbf{h_{2}}$       & x-vector \\ \Xhline{2.5\arrayrulewidth}
speaker ($k=200$)   & 92.20 & 65.10         & 87.90    \\
noise ($k=4$)     & 0.10            & 0.70  & 0.10      \\ 
mic placement ($k=12$)      & 0.10           & 2.00 & 1.00     \\ \Xhline{2.5\arrayrulewidth}
\end{tabular}
\end{table}

\begin{table}[!t]
\caption{Diarization with oracle speech segment boundaries and known number of speakers (AMI)}
\label{table:sd}
\setlength{\tabcolsep}{8.5pt}
\centering
\begin{tabular}{c|ccc}
\Xhline{2.5\arrayrulewidth}
System        & Baseline 1 & Baseline 2 & $\mathbf{h_{1}}$   \\ \Xhline{2.5\arrayrulewidth} 
Avg. DER (\%) & 11.91      & 11.51      & \textbf{7.28}                                                     \\ \Xhline{2.5\arrayrulewidth}             
\end{tabular}
\end{table}

\section{Conclusions}
\label{sec:conclusions}
We present an adversarial invariance approach to obtain speaker embeddings robust to various sources of acoustic variability present in speech signals. The embeddings are learnt by disentangling speaker-related information from all other factors without supervised information about the acoustic conditions. We evaluate these embeddings for various tasks such as speaker verification, clustering and diarization. Experimental results suggest that our method is able to produce robust speaker embeddings in a variety of challenging acoustic scenarios.
In the future, we will focus on obtaining speaker representations using low-level audio features such as spectrograms, while further improving their robustness in other challenging acoustic conditions.

\bibliographystyle{IEEEbib}
\bibliography{strings,refs}

\end{document}